# School-based malaria chemoprevention as a cost-effective approach to improve cognitive and educational outcomes: a meta-analysis

Noam Angrist[1,2,3*], Matthew C. H. Jukes[4], Sian Clarke[5], R. Matthew Chico[5], Charles Opondo[6], Donald Bundy[7], Lauren M. Cohee[8*]

[1]Youth Impact, Gaborone, Botswana

[2]Department of Economics and Blavatnik School of Government, University of Oxford, Oxford, UK

[3]The World Bank, Washington, DC, USA.

[4]RTI International, London, UK

[5]Department of Disease Control, Faculty of Infectious and Tropical Diseases, London School of Hygiene & Tropical Medicine, London, UK.

[6]Department of Medical Statistics, Faculty of Epidemiology & Population Health, London School of Hygiene & Tropical Medicine, London, UK.

[7]Research Consortium for School Health and Nutrition, London School of Tropical Hygiene and Medicine, University of London, London, UK

[8]Center for Vaccine Development and Global Health, University of Maryland School of Medicine, Baltimore, USA.

*Corresponding authors:
Angrist: nangrist@youth-impact.org
Cohee: lcohee@som.umaryland.edu

**ABSTRACT**

There is limited evidence of health interventions' impact on cognitive function and educational outcomes. We build on two prior systematic reviews to conduct a meta-analysis, exploring the effects of one of the most consequential health interventions, malaria chemoprevention, on education outcomes. We pool data from nine study treatment groups (N=4,075) and outcomes across four countries. We find evidence of a positive effect (Cohen's d = 0.12, 95% CI [0.08, 0.16]) on student cognitive function, achieved at low cost. These results show that malaria chemoprevention can be highly cost effective in improving some cognitive skills, such as sustained attention. Moreover, we conduct simulations using a new common metric – learning-adjusted years of development – to compare cost-effectiveness across diverse interventions. While we might expect that traditional education interventions provide an immediate learning gain, health interventions such as malaria prevention can have surprisingly cost-effective education benefits, enabling children to achieve their full human capital potential.

**SIGNIFICANCE STATEMENT**

While health and education are interlinked, the two sectors often operate in silos. Researchers rarely assess the impact of health interventions on education outcomes, leaving complementarities unknown and undervalued. This study contributes evidence on the effectiveness of a consequential health intervention in low- and middle-income countries – malaria chemoprevention – on cognitive and educational outcomes. We conduct a meta-analysis of school-based malaria chemoprevention randomized controlled trials comparing impacts across common education metrics. Results show that health interventions such as malaria prevention can have surprisingly cost-effective education benefits, warranting serious consideration in the toolkit of approaches to improve children's education. More broadly, results reveal the value of evaluating impacts across sectors.

# INTRODUCTION

Despite evidence that health and education are intrinsically interlinked – healthy individuals more likely to acquire and complete education (Andersen et al. 2021), and educated individuals more likely to lead healthy lives (De Neve et al. 2015) – the two sectors often operate in silos. Indeed, researchers in both fields rarely assess the impact of health interventions on education outcomes, leaving complementarities unknown and undervalued (Kruk et al. 2022). For example, a landmark review of malaria control on *Plasmodium falciparum* in Africa between 2000 and 2015 reported that insecticide-treated nets, the most widespread intervention avert to malaria cases, did not explore the impact of interventions on education (Bhatt et al. 2015). Neither did a prominent Cochrane review of insecticide-treated nets published the following year (Pryce et al. 2018).

However, multi-sectoral outcomes have revealed surprising findings when investigated. For example, school-based deworming, primarily a health intervention, has shown cost-effective effects on education outcomes, such as schooling attendance in Kenya (Kremer and Miguel 2004), with long-run effects on earnings (Hamory et al. 2021). Similarly, school-feeding interventions, an important category of school-based health interventions with over 400 million children receiving school meals worldwide, have shown positive effects on schooling and learning (Kazianga et al. 2012; Aurino et al. 2020).

In this paper, we explore cognitive and education outcomes attributable to the prevention of malaria, one of the most consequential global health challenges today. Observational studies have associated malaria infection among schoolchildren to lower cognitive function and performance in school (Fernando 2003a, Fernando 2003b, Fernando 2003, Simwaka 2009, Thuilliez 2010, Vitor-Silva 2009, Nankabirwa 2013, Vorasan 2016, MacNab 2016, Brasil 2017, Orish 2018). A follow up of early childhood malaria chemoprophylaxis found a long-term impact on cognitive function and education achievement (Jukes et al, 2006). Yet evidence from randomized-trials to explore the impact of malaria prevention on education outcomes is more limited. Examining links between school-based malaria prevention and education is timely: school health has been identified as a key priority for achieving the Sustainable Development Goals (Bundy et al. 2017) and the WHO recently included school-based chemoprevention in malaria control guidance (https://app.magicapp.org/#/guideline/6832).

In this study, we pooled all available data from randomized trials of school-based malaria chemoprevention that captured effects on education outcomes and cognitive function. We conducted one of the first cost-effectiveness comparative analyses with traditional education interventions. We focused on school-based interventions since schools provide a cost-effective and scalable delivery platform, and the potential to affect education outcomes is high (Fig. 1). Specifically, we assessed the impact and cost-effectiveness of reducing *Plasmodium* infections, which cause malaria disease, in schoolchildren on cognition and education outcomes capturing schooling and learning. We considered performance in tests of sustained attention - the ability to maintain focus over a long period of time (Fortenbaugh et al., 2017) - and working memory - the ability to keep information in mind and use it (Zelazo et al., 2017) - as cognitive skills with the potential to translate into improved learning.

An estimated 200 million school-age children are at risk each year of *P. falciparum* infection in sub-Saharan Africa with one-half of schoolchildren infected at any given time in many endemic areas (Bundy et al. 2017,

Yapi et al. 2014, Mathanga et al. 2015, Pinchoff et al. 2016, Clarke et al. 2017, Were et al. 2018). *Plasmodium falciparum* infection rarely results in severe illness or death in this age group, but commonly causes acute clinical malaria and chronic infection. These chronic infections are often termed asymptomatic, but they are associated with anemia, malaise, and feeling unwell (Chen et al Plos Med. 2016). The adverse effects of *P. falciparum* infection that impact the education of schoolchildren occur through at least three causal pathways: (a) school absences leading to decreased opportunities for learning, (b) impaired cognitive function limiting short-term ability to benefit from learning opportunities, and (c) impaired cognitive development limiting longer-term benefit from learning opportunities (Fig. 1). To this end, cost-effective interventions to improve health and education outcomes might enable children to achieve their full human capital potential.

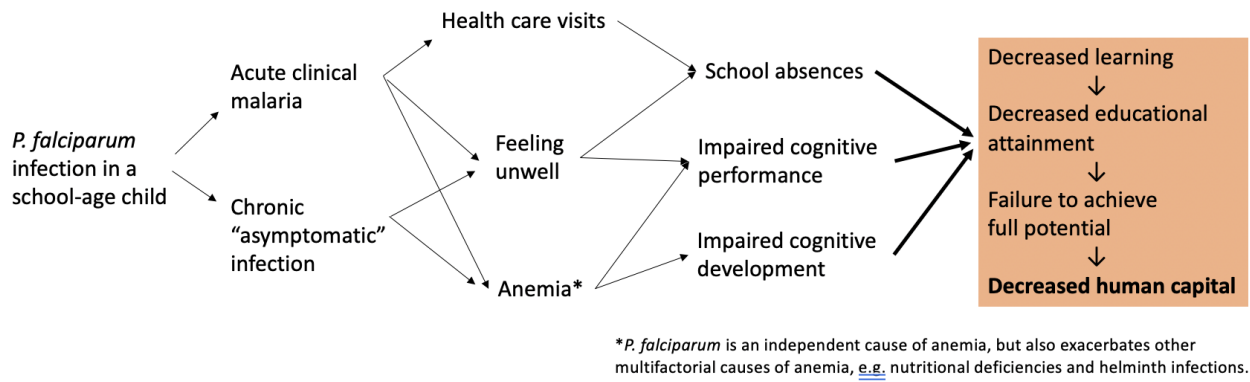

**Fig. 1.** Path representation of the causal effect of *Plasmodium falciparum* infection on human capital

**RESULTS**

**Study description and design**. We conducted a random-effects meta-analysis of six randomized controlled trials of school-based malaria chemoprevention (N=4,075) involving nine unique study arms. We pooled chemoprevention studies that reduce *Plasmodium* infection and/or anemia, representing the most plausible pathway to educational impact (Fig. 1). We build on an earlier systematic review (Cohee et al. 2021), adding analysis of intervention effects on a more comprehensive set of education outcomes.

We calculated standardized education outcomes across studies to facilitate pooling data. Our primary analysis was conducted with standardized education outcomes, including cognitive skills and educational achievement. In addition, we conducted secondary analysis with outcomes expressed in terms of an education measure increasingly used by organizations such as the World Bank called the Learning-Adjusted Years of Schooling (LAYS) measure (Filmer et al. 2020; Angrist et al. 2020). This measure combines schooling and learning into a single composite measure. LAYS are the education analogy to DALYs in the health sector (as estimated in the annual Global Burden of Disease study), enabling value-for-money comparisons across a range of outcomes. LAYS can be interpreted as a high-quality year of schooling – that is, schooling which results in substantial learning – according to global benchmarks. The measure has gained prominence in education, such as forming the education pillar of the World Bank Human Capital Index. In the context of our meta-analysis, since we included cognitive skills that could translate into later educational gains, we interpret this measure as "Learning-Adjusted Years of Development (LAYD)", referring to learning accumulated as children develop over time, rather than only learning gained through schooling.

We grouped trials into two main categories by outcome: (a) cognitive skills such as sustained attention, and (b) scores in tests of educational achievement, such as literacy and numeracy. Sustained attention and working memory are cognitive skills which are most responsive to health improvements in the short term. If these improvements in cognitive function are maintained over long periods, they can result in improved learning, such as acquiring literacy and numeracy skills. We also include a third category for studies with a distinct epidemiology, specifically, trials in areas where *Plasmodium vivax* infections are also prevalent, as clearing these infections, which are often recurrent, might make malaria prevention interventions in schools particularly effective.

**Effect sizes on educational and cognitive outcomes.** When pooling studies (Fig. 2), we found strong evidence of a positive effect size of Cohen's d = 0.163 (95% CI [0.079, 0.248], z = 3.788, P < 0.001). For cognitive skills outcomes, we also found a strong effect size, d = 0.122 (95% CI [0.085, 0.160], z = 6.400, P < 0.001). However, for literacy and numeracy outcomes, we found no evidence of an effect, d = 0.008 (95% CI [-0.125, 0.141], z = 0.116, P = 0.908). For the interventions tested in a setting with distinct epidemiology, Sri Lanka in 2006, however, we found strong evidence of a large effect size on literacy and numeracy outcomes of d = 0.399 (95% CI [0.342, 0.456], z = 13.666 P < 0.001). Within each subgroup, we found little heterogeneity; $I^2$ is 0 in each of the three sub groupings, suggesting that our findings are generalizable within these categories, with effects statistically similar across studies and contexts.

**Simulating effects using global human capital measures.** We conducted a simulation analysis using the LAYD concept (Figure 3). In conducting this calculation, we considered performance in tests of sustained attention and working memory as cognitive skills which can translate into improved learning. For the pooled sample we found an overall effect of 0.204 (95% CI [0.098, 0.310], z = 3.788, P < 0.001). For cognitive skills, we estimate an effect of 0.153 (95% CI [0.106, 0.200], z = 6.400, P < 0.001). For literacy and numeracy, we found an effect of 0.010 that is not statistically different from zero (95% CI [-0.157, 0.176], z = 0.116, P = 0.908). For interventions with a distinct epidemiology, we found an effect of 0.499 on literacy and numeracy outcomes (95% CI [0.427, 0.570], z = 13.666, P < 0.001). In terms of heterogeneity, we found the same results as when we performed the analysis on the unscaled effects, with little heterogeneity indicated by an $I^2$ of 0 for each of the three sub-groupings.

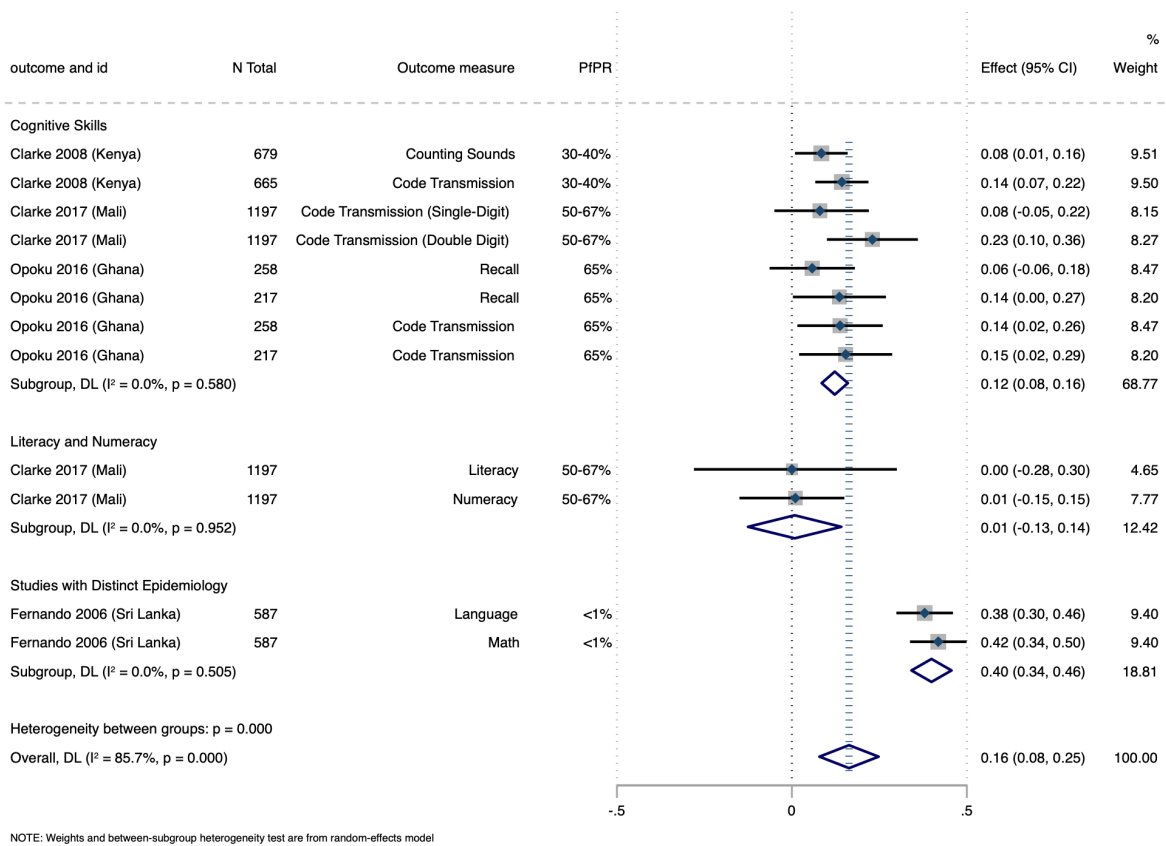

**Fig. 2.** Forest plot of effects of malaria chemoprevention on cognitive skills using standard deviations

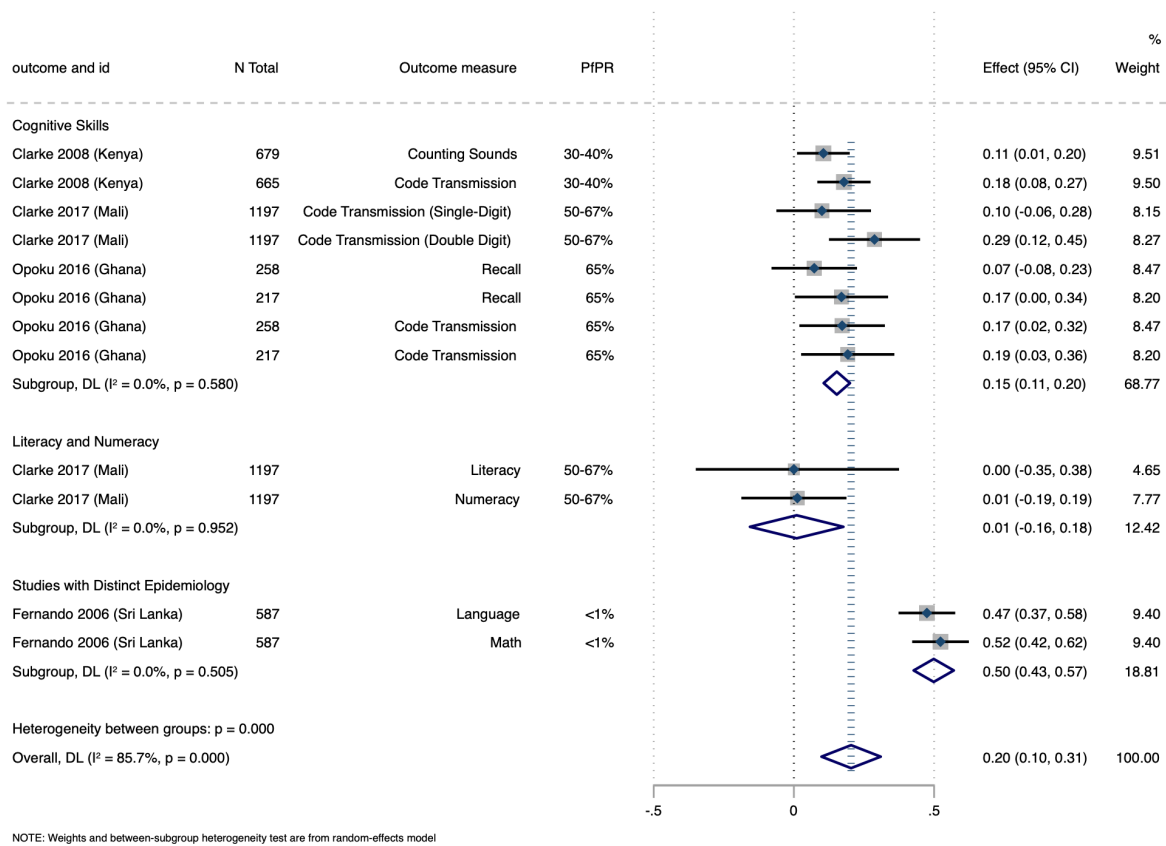

**Fig. 3.** Forest plot of effects of malaria chemoprevention on learning-adjusted years of development

**Cost-effectiveness comparative analysis.** We next compared the cost-effectiveness of interventions focused on malaria chemoprevention with other well-known programs in education (Fig. 4). A recent review compiled results across 150 impact evaluations in education, including categories such as teacher incentives, general skills teacher training, scholarships, and education technology, among others (Angrist et al. 2020). One finding from this prior review was that many education programs and policies which are popular are not as effective as typically believed to be, such as general skills teacher training and simply providing more inputs into classes, e.g. additional textbooks and computers. These findings prompt the need to identify additional types of interventions which can cost-effectively improve education outcomes.

We compared intervention types (Fig. 4) which have shown effectiveness in education alongside school-based malaria chemoprevention. To move from effectiveness to cost-effectiveness analysis we included costs in studies for which cost information was available. Clarke et al. (2008) find a cost per child of $1.88 per child treated per year; Clarke et al. (2017) found a cost per child per year of $2.72. We assumed these estimates apply across studies where cost data was not collected, since the cost of delivering interventions can be similar across similar intervention typologies.

Results (Fig. 4) revealed that malaria chemoprevention, on average, performs well relative to many cost-effective programs in education, such as teacher incentive reforms and merit scholarships. In an analysis which included all school-based malaria chemoprevention studies, the gains are equivalent to a median of five learning-adjusted years of development per $100. In a second analysis, which considered only studies capturing effects on cognition outcomes, which had the largest effects in the meta-analysis, this estimate yields a median of eight learning-adjusted years of development per $100.

Overall, in terms of comparative cost-effectiveness with more traditional education interventions, malaria chemoprevention ranked second on average, higher than popular education programs and policies, such as teachers incentive reforms. These results reveal the striking potential for health interventions to have efficient returns to some cognitive and education outcomes. These results are driven by a combination of moderate effect sizes on par with average effective education interventions; for example, we found an effect of 0.12 standard deviations on cognition, and the average education effectiveness is approximately 0.10 standard deviations (Evans and Yuan 2022), but with substantially lower costs. However, the direct comparison with educational investments is imperfect, since education interventions most often measure literacy and numeracy outcomes rather than emergent cognitive skills.

**Fig. 4.** Cost-effectiveness of malaria chemoprevention compared to education interventions

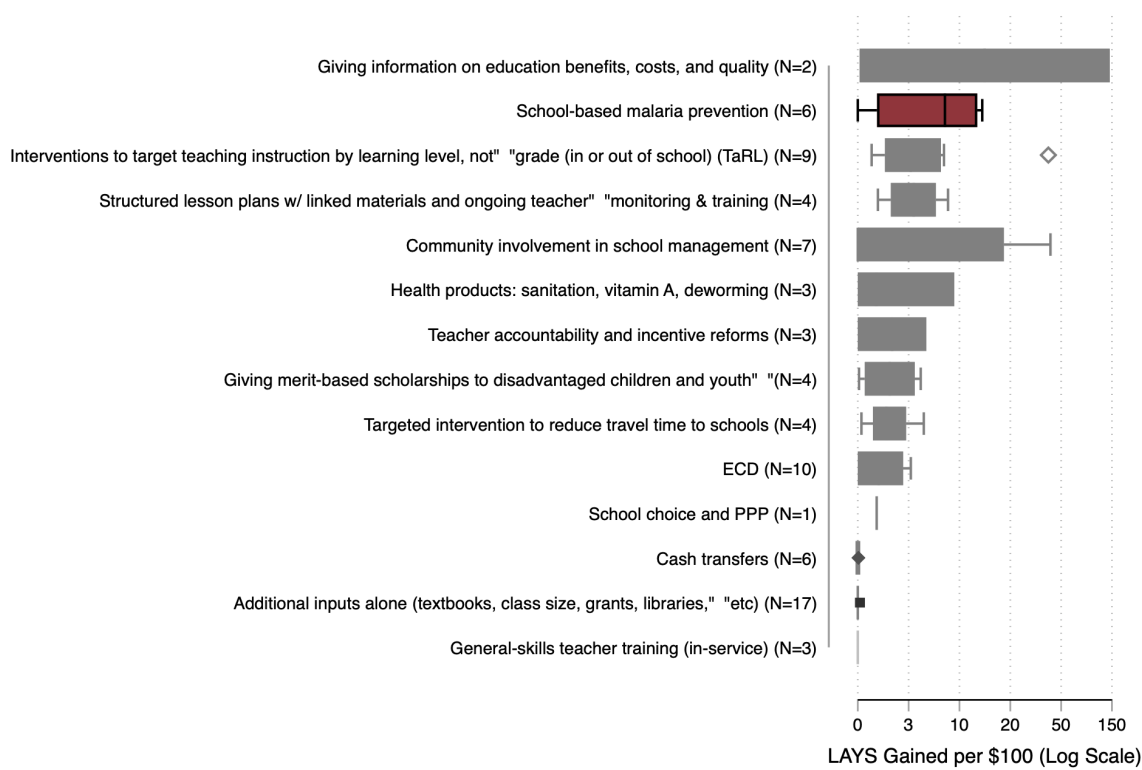

*Notes*: This figure is adapted from Angrist et al. (2020) which includes over 150 impact evaluations grouped into various categories. These categories range from 'additional inputs (such as grants to schools, and computers)' to 'early childhood development.' We overlay our results in terms of school-based malaria chemoprevention in red, revealing high cost-effectiveness.

**Robustness checks and publication bias.** Our results are robust to incorporating additional yet unpublished results (Appendix Figures 1 and 2). We found evidence of positive and statistically significant effects (Cohen's d = 0.11, 95% CI [0.04, 0.18]), although $I^2$ within categories are now higher, suggesting more heterogeneity.

We also performed checks for publication bias, including a funnel plot analysis as well as checking for unusually clustered t-statistics above statistical significance thresholds (Appendix Figures 3, 4, 5 and 6). Inspection of our funnel plot suggests a symmetric distribution of effect sizes and their corresponding standard errors (Appendix Figures 3 and 4). An Egger's test on the sample without unpublished interventions indicates no evidence of small study effects (b = -0.007, 95% CI [-0.809, 0.794], t(12) = -0.02, P = 0.984). On the sample including unpublished results, the Egger's test also indicates no evidence of small study effects (b = -0.034, 95% CI [-0.703, 0.633], t(16) = -0.11, P = 0.913). Moreover, histograms for the distributions of t-statistics show limited concentration right above the 1.96 cutoff, suggesting limited publication bias (Appendix Figures 5 and 6).

**DISCUSSION AND CONCLUSION**

This paper aggregates evidence from school-based randomized controlled trials, showing malaria chemoprevention can be cost effective in terms of improving certain education outcomes, such as cognitive skills. While we might expect that traditional education interventions provide an immediate learning gain, health interventions, such as school meals, malaria prevention, and deworming, might have surprisingly cost-effective education benefits. Moreover, these results highlight the importance of measuring education effects for health interventions. Without measurement across sectors, the more comprehensive impact of interventions goes uncounted, unnoticed, and undervalued.

While our findings highlight the potential for school-based malaria chemoprevention to improve education, there are important limitations to our analyses. We do not consider other forms of malaria prevention, such as bednets (Cohen and Dupas 2008) in this analysis, which might also have educational benefits. We further assume that improvements in health occur along the causal pathway to improvements in education and therefore included in our analysis only trials in which the intervention was effective at improving health outcomes – that is, decreasing *P. falciparum* infection and/or anemia. Thus, our findings apply when interventions successfully improve health. However, because the interventions included in our analysis were relatively short-term, ranging from a one-time treatment to, at most, two years of chemoprevention, our results may underestimate the benefits as it may require longer periods of intervention for improved cognitive function and catch-up development to translate into increased educational attainment. That said, our findings are supported by recent population-level malaria elimination efforts including mass anti-malarial drug administration which were associated with increased school-attendance and improved test scores across all subjects (Cirera 2022).

A caveat to our findings is that we compare the cost-effectiveness of interventions using various outcomes. We use cognitive function as an outcome for malaria interventions and make comparisons with educational impacts on achievement tests. We analyze these cognitive and educational outcomes together for illustrative purposes, although further work is needed to understand how enhanced cognitive skills following malaria chemoprophylaxis translate into improved learning. For now, we argue that the high cost-effectiveness of malaria interventions to improve cognitive function warrant their serious consideration in the toolkit of approaches to improve children's education.

Our findings support the notion of synergies between the health and education sectors to increase human capital. By recognizing the multi-sectoral impacts of health interventions, we may also be better able to overcome the challenges of implementing and financing such interventions. Future trials and implementation of school-based malaria prevention should measure both health and education outcomes, ideally longer-term education outcomes using standard measurements to facilitate further comparative analyses. Combining school-based malaria chemoprevention with other health interventions, such as nutrition and deworming, through integrated school health platforms could further decrease costs and increase benefits. Evaluating the multiplicative impacts of combined interventions may reveal additional complementarities and cost-effective approaches to unlock children's full human capital potential.

**METHODS**

**Eligibility Criteria**: We included studies from a systematic review that included randomized studies that assessed the effect of antimalarial treatment among asymptomatic school-aged children aged 5–15 years in sub-Saharan Africa and were published between Jan 1, 1990, and Dec 14, 2018 (Cohee et al. 2021). No further studies meeting these criteria have been published through Mar 1, 2023. One study that did not meet inclusion criteria for the prior review because it took place outside of sub-Saharan Africa was included as it was a seminal study generating interest in this area. Because intervention impact on cognition and education is mediated by parasite clearance and/or improvement in hemoglobin, only studies in which the intervention decreased *P. falciparum* infection and/or anemia are included in this meta-analysis. We include published studies only in the main analysis, with unpublished studies included in corollary analysis in the appendix.

**Intervention Details:** All trials evaluated school-based administration of anti-malarial drugs to students as chemoprevention to clear existing infections and provide a period of chemoprophylaxis, the drug administered and the number of doses, frequency and interval between doses varied between the trials.

**Outcomes**: Results from the most comprehensive and standardized set of education outcomes to date are included in this meta-analysis, including cognitive skills measured by sustained attention and working memory as well as literacy and numeracy. Cognitive skills such as sustained attention outcomes include counting sounds and code transmission tests and working memory was measured by three object recall assessed after ten minutes. Literacy and numeracy outcomes include routine examinations or tests developed to measure learning of standard classroom curricula.

**Data**: We standardize outcomes across studies as well as including cost data to inform cost-effectiveness analysis. We standardize outcomes relative to the standard deviation in the control group. Our primary analysis is conducted with these standardized outcomes yielding Cohen d's effect sizes. In addition, we conduct secondary analysis with outcomes expressed in terms of a new measure in education called the learning-adjusted years of schooling (LAYS) measure (Filmer et al. 2020; Angrist et al. 2020). LAYS are a composite measure which can be interpreted as a high-quality year of schooling in which students learn a substantial amount, and are in many ways analogous to DALYs in the health sector, enabling value-for-money comparison across a range of outcomes. The measure combines global schooling data from UNESCO and globally harmonized learning outcomes (Angrist et al. 2021; Kraay 2019). In addition to use in the World Bank Human Capital Index, the LAYS metric is used to review global evidence by the Global Education Evidence Advisory Panel (GEEAP).

In terms of calculating LAYS, literacy and numeracy are primarily learned in school, enabling direct and clear interpretation. Cognitive skills such as sustained attention, however, develop in tandem with changes in brain function and in response to experiences in and out of the classroom. To this end, the term learning-adjusted years of schooling should be interpreted with this caveat since we interpret learning trajectories in terms of early years, where increasingly many children, but not all, are in some form of school. In the context of our meta-analysis, since we include both cognitive and educational measures, we interpret this measure as "Learning-Adjusted Years of Development (LAYD)", referring to learning accumulated as children develop over time, rather than only learning gained through schooling.

**Statistical Methods:** We conduct a random-effects meta-analysis using *Stata* software. We also segment results by two major outcome types: (a) sustained attention and working memory outcomes and (b) literacy and numeracy outcomes. Sustained attention and working memory are early cognition skills and most plausibly linked to immediate health improvements which in turn can improve cognitive functioning, whereas literacy and numeracy are likely to be gained due to compound over the long run. We also include a third category for studies with a distinct epidemiology, specifically, trials in areas where *Plasmodium vivax* is prevalent in addition to *P. falciparum,* where recurrent infections might make malaria prevention interventions particularly effective. We calculate standard effect sizes, such as Cohen's d-statistics, 95% confidence intervals, as well as $I^2$ statistics which quantify heterogeneity.

**Supplementary Information**: Figures S1 to S6

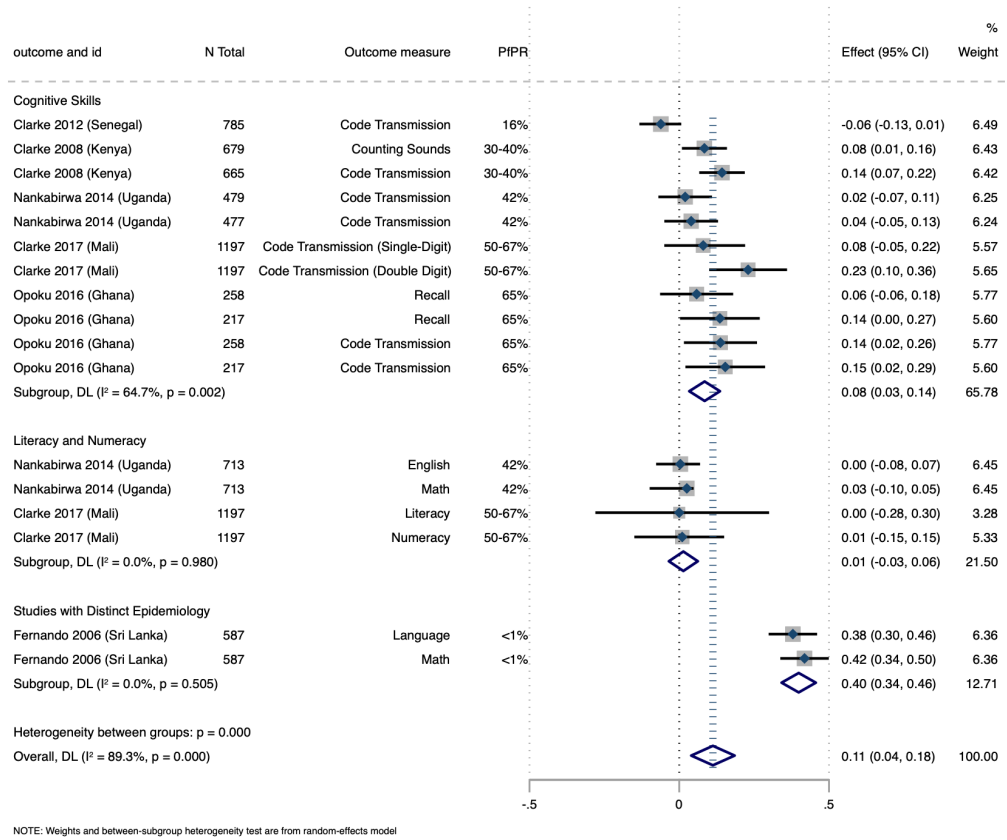

**Fig. S1.** Forest plot of effects of malaria chemoprevention on learning outcomes using standard deviations, including unpublished studies.

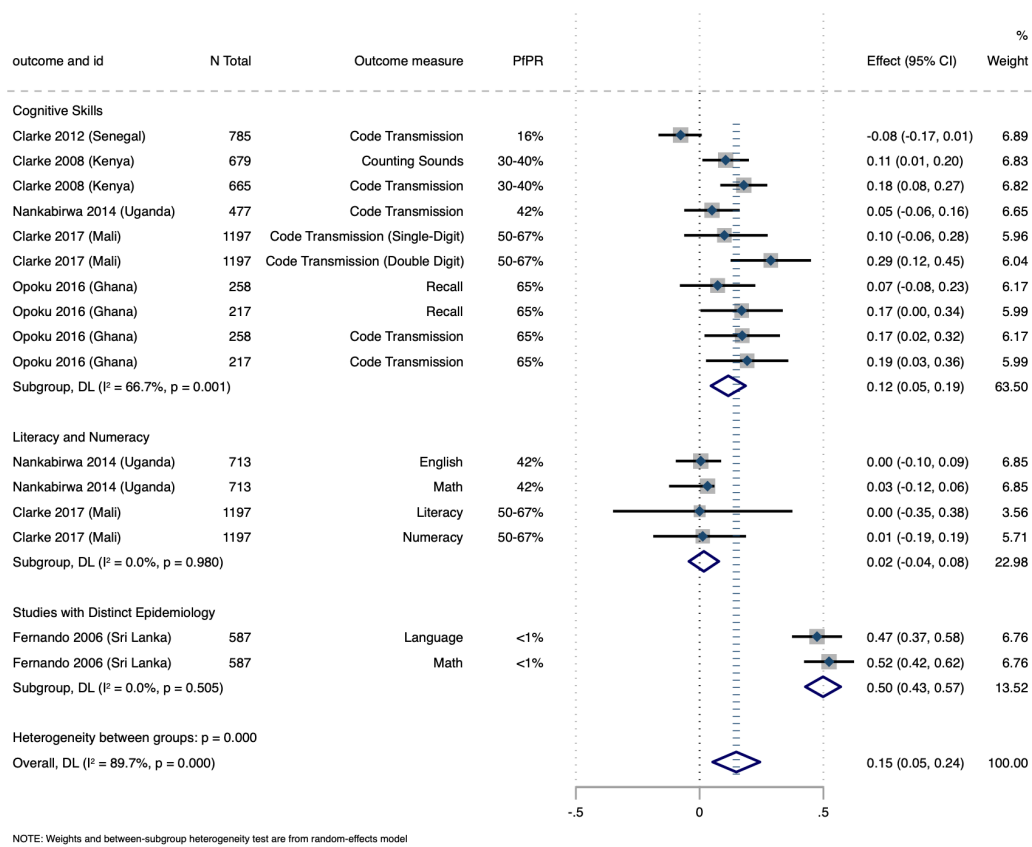

**Fig. S2.** Forest plot of effects of malaria chemoprevention on learning outcomes using LAYS including unpublished studies.

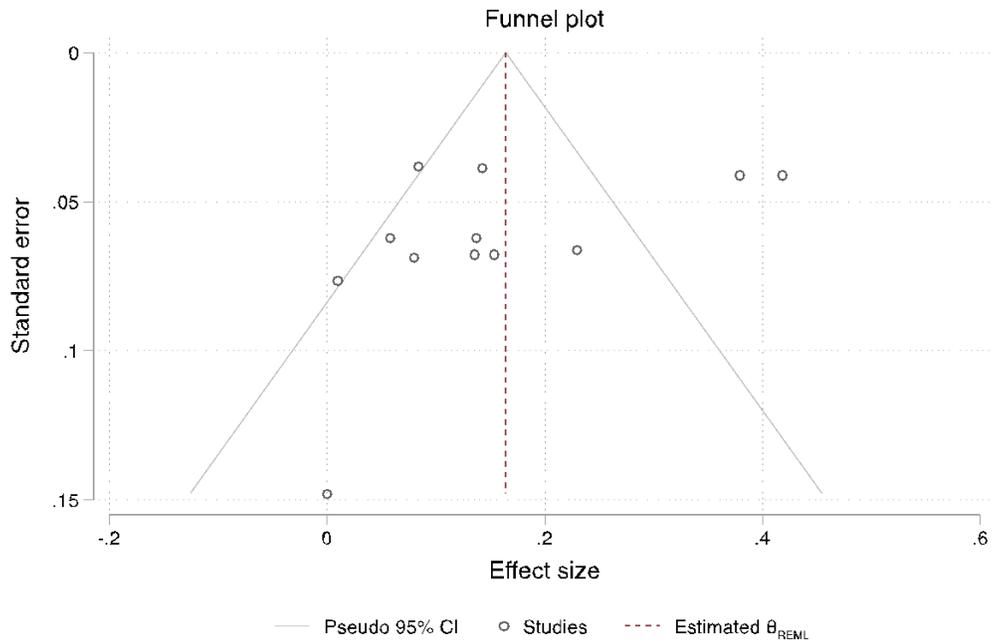

**Fig. S3.** Funnel plot for effect sizes of malaria chemoprevention on learning outcomes, including published studies.

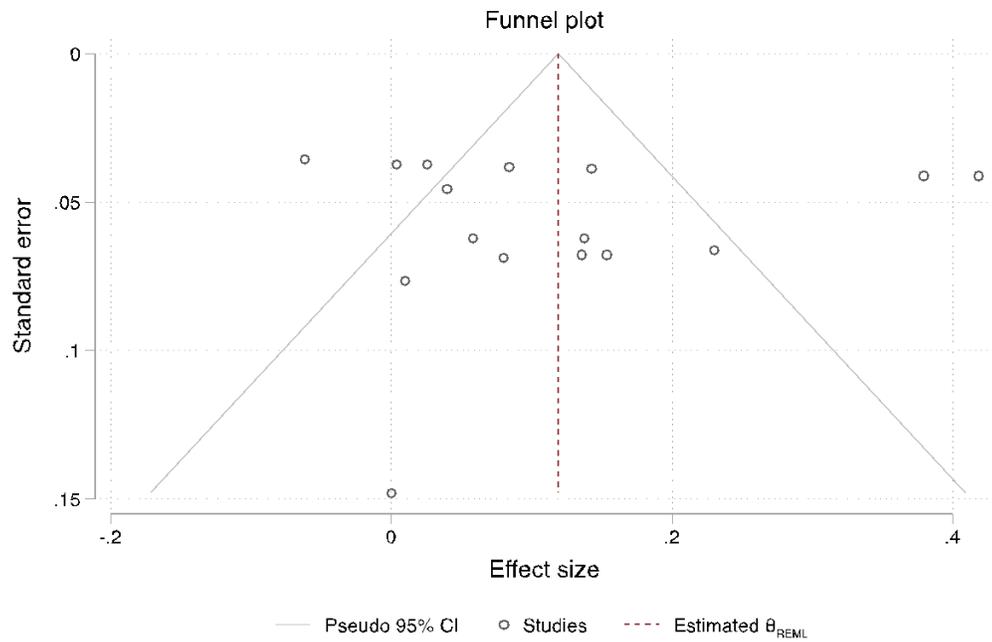

**Fig. S4.** Funnel plot for effect sizes of malaria chemoprevention on learning outcomes, including unpublished studies.

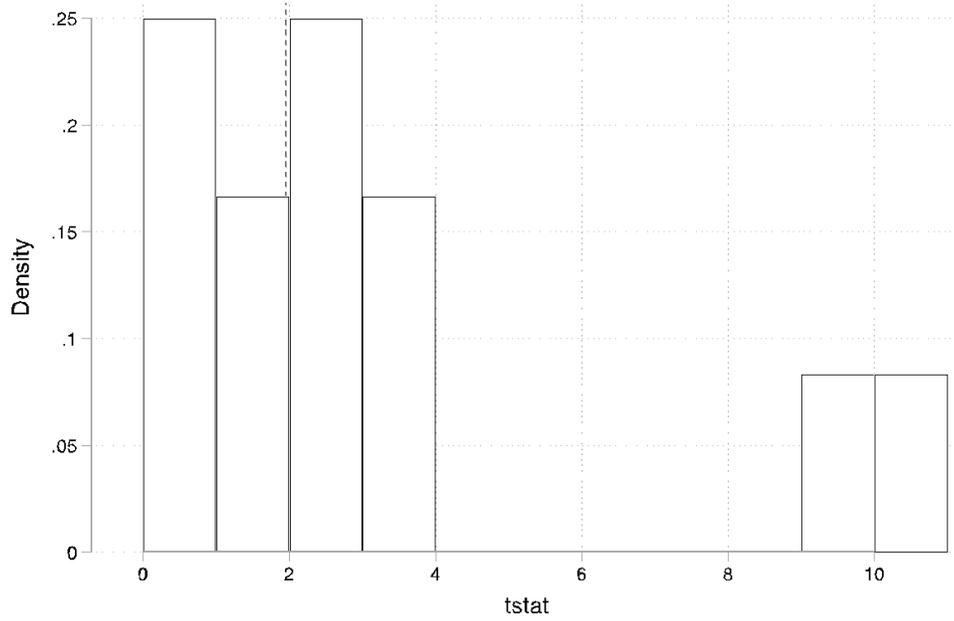

**Fig. S5.** Distribution of t-statistics of estimates of the effect of malaria chemoprevention on learning outcomes, including published studies.

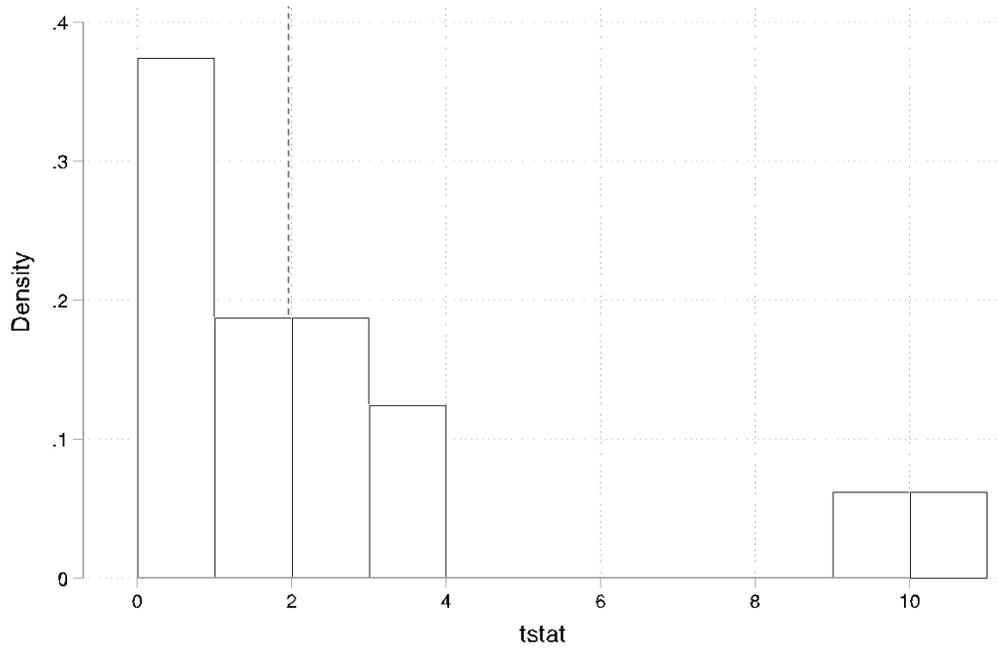

**Fig. S6.** Distribution of t-statistics of estimates of the effect of malaria chemoprevention on learning outcomes, including unpublished studies.